\documentclass[12pt]{article}  
\usepackage{epsfig} 
\usepackage{graphics} 
\usepackage{rotate} 

\def \bg{\bigskip} 
\def \no{\noindent} 
\begin{document}
% \Large  
%\normalsize
  \small

 %\setcounter{page}{1}
%\bg 
 % Balance,  June,   3, 2002
% .

\bg

 \bg

\bg

{\centerline{\large {\bf Is the Energy Balance Paradox Solved? }} } 

%\bg

 \bg

\bg
 
{\centerline  {\bf Amos Harpaz(1)  and Noam Soker(2) }} 

\bg

\no 1. Institute of Theoretical Physics, Technion, Haifa, ISRAEL

\no 2. Department of Physics,  
Oranim, Tivon 36006, ISRAEL  

 \bg

 \no  phr89ah@tx.technion.ac.il 

\no soker@physics.technion.ac.il

\bg

\bg

 \no key words: Radiation Reaction Force, Curved Electric Field

\bg

\no {\bf   Abstract  }   

The question of what is the source of the energy carried by radiation 
is discussed.  The case of a hyperbolic motion is analyzed, describing 
the solutions suggested in the past for the ``energy balance paradox".  
 The solution  to the paradox is found in considering the stress force 
created in the curved electric field of the accelerated charge, that 
acts as a reaction force.  The work performed by the external force  
   to overcome this stress force is the source of the energy carried by  
the radiation.  This stress force is the 
spatial component of the four-vector called ``Schott term", that appears 
in  Abraham four-vector.   This novel approach solves the ``energy balace oaradox".  

 \bg

 \bg

Pacs: 41.60-m,  03.50-DE
 \bg

  \vfil

  \eject

\bg

\no {\bf 1.  Introduction} 
\bg

The Energy Balance Paradox concerns the question of - what is the source 
of the energy carried by radiation.  This problem is complicated and demands a careful 
analysis of all the factors relevant to the process.   Hence, we shall concentrate 
here on the most simple case of creation of radiation - a charge accelerated uniformly 
in its own system of reference, in which case its motion is described in a four 
dimensional free space as a hyperbolic motion [1].  

One expects that the energy carried by the radiation is created by a work done by 
the external force against a certain force that exists in the system, where this 
work should be performed in addition to the work done in creating the kinetic energy 
of the charged particle.  In many publications this problem is treated by using 
the concept of ``radiation reaction force", where it is assumed that a reaction 
force created by the radiation is the force that the external force should overcome, 
and the work done against the  reaction force  
is the source of the energy carried by the radiation.  However, in the simple case 
of the hyperbolic motion this approach meets difficulties      
that were named ``The Energy Balance Paradox", as  no such force 
exists in this  motion.  

The difficulties have two sources:  one - when the accelerated charge moves with 
low (zero) velocity, its radiation is symmetric with respect to the
 plane which is perpendicular to the direction of motion.  
Thus the radiation does not impart any counter momentum to the accelerated charge, 
and no radiation reaction force exists.  
The second source of the difficulties is that usually, it is considered that in 
the equation of motion of the accelerated charge, a term called Abraham four 
vector is  considered as representing the radiation reaction force.  It 
comes out that in a hyperbolic motion, this vector vanishes.  This suits the fact that 
in this case no radiation reaction force exists, but it leaves us with the paradox.    

   In the present work we shall 
analyze the paradox, and the solutions given to it.  We shall also show that the 
solution to the paradox is found when a reaction force is identified, which is 
not the ``radiation reaction force", but rather a stress force that exists in 
the curved electric field of the accelerated charge.  The work done by the 
external force to overcome the stress force, is the source of the energy carried 
by the radiation.  
 
\bg

\no {\bf  2. The Problem.}  

\bg

The formula for the angular distribution of the radiation power is [2]: 

$${dP\over d\Omega} = {e^2 a^2\over 4\pi c^3}{\sin^2 \theta\over 
 (1 - \beta \cos \theta)^5}         \eqno(1) $$
\no where $e$ is the accelerated charge, $a$ is the acceleration,  $c$ is the 
speed of light, and $\theta$ is the angle measured from the direction of 
motion.  Integrating eq. 1 over the angles, yields: 

$$ P = {2\over 3}{e^2a^2\over c^3(1-\beta^2)^3} = 
 {2\over 3}{e^2 (\gamma^3 a)^2\over c^3}    \eqno(2)  $$ 

\no where for $\beta \rightarrow 0 \ \ (\gamma \rightarrow 1)$, yields Larmor  
formula for the power carried by the radiation: 

$$ P = {2\over 3}{e^2 a^2\over c^3}    \eqno(3)   $$ 

In Fig. 1 we plot the angular distribution of the radiation for several values of $\beta$. 
It is clearly observed that for low velocities ($\beta \le 0.01$), no counter momentum 
is imparted by the radiation to the charge, and no radiation reaction exists.  

\begin{figure}
\centering\epsfig{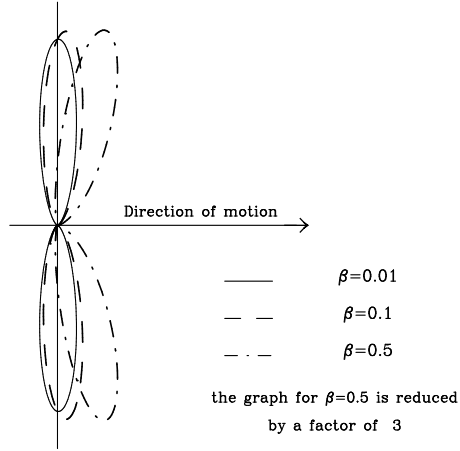}
\vskip 0.2cm
\caption{\small The angular distribution of the radiation}
\end{figure}

The Dirac-Lorentz equation (or LAD as called by Rohrlich[3]) for a charged particle is: 

$$ ma^{\mu} = F^{\mu}_{ext}  + \Gamma^{\mu} = F^{\mu}_{ext} +
 {2 e^2\over 3c^3}\left(\dot{a}^{\mu} - {1\over c^2} a^{\lambda} a_{\lambda} 
 v^{\mu}\right) ,      \eqno(4)  $$

\no where $m$ is the particle mass,  $v^{\mu}$ and $a^{\mu}$ are the velocity 4-vector  and the 
 acceleration 4-vector, respectively, and $F^{\mu}_{ext}$ is the external force 4-vector 
 that drives the particle.  $\Gamma^{\mu}$ is  Abraham  4-vector, which was usually 
considered as representing  the radiation reaction force.   
          
%\no where  $v^{\mu}$ is the 4-vector velocity of the charge:  $v^{\mu} = \left(
%\matrix{\gamma c  \cr  \gamma \vec{v}}  \right)$.  

The first term in $\Gamma^{\mu}$, (${2\over 3}{e^2 \dot{a}^{\mu}\over c^3}$), is 
called the ``Schott term".   In a hyperbolic motion, $\dot{a}^{\mu} = 
{1\over c^2}a^{\lambda}a_{\lambda} v^{\mu}$, and $\Gamma^{\mu}$ vanishes.   This 
is in accord with the nonexistence of a radiation reaction force in such a motion as 
shown in Fig. 1 (concerning the conjencture that   $\Gamma^{\mu}$ represnts the 
radiation reaction force).    In the present work we follow Rohrlich[3], who argues that the 
reaction force responsible for the creation of the radiation should be included in Schott term, 
thus showing that the vanishing of $\Gamma^{\mu}$ shows that the power carried by the 
radiation, which is represented in the second term of $\Gamma^{\mu}$ is created by the force 
represented in Schott term.  However, this force is not a radiation reaction force, but rather 
a reaction force created by the stress force the exists in the curved electric field of 
the accelerated charge.  This  reaction force should be 
overcome by the accelerating external force, and the work performed against this 
force is the source of the energy carried by the radiation (see section 4). 

The presence of $\dot{a}$ (the third time derivative of the position) in the 
equation of motion (eq.4), demands a third  initial condition for the solution 
of the equation of motion of an electric charge
 (the initial acceleration).  In section 4 we find that 
a stress force, $f_s$, which exists in the curved electric field of the accelerated 
charge, is proportional to the acceleration.   Thus, the third initial condition is needed 
to complement the picture of all the factors involved in such a motion.  It is 
also found that this force, $f_s$, is responsible for the creation 
of the radiation. 
  
\bg

\no {\bf  3. Solutions suggested }  

 \bg

One of the solutions was [4], that there exists a charged plane, 
 whose charge is equal and opposite in sign to the accelerated 
 charge, and it recesses with the speed of light in a direction 
opposite  to the direction of the acceleration.  The interaction between 
 this charged plane and the accelerated charge creates the energy 
carried by the radiation.  
Another suggestion was [5, (eq. 4.9)],  that in such a case the 
energy radiated is supplied from the self-energy of the charge.  
Evidently, these suggestions are far from being satisfactory. 
Some people deduced that a uniformly accelerated charge does not radiate 
(see Singal[6]).  
 It should be noted that the idea suggested in  [4]  resolves another 
difficulty concerned with this topic, which is the existence of a single 
electric charge.   As we assume that the matter in the universe is 
neutral, the existence of a solitary charge is a local phenomenon, whose 
validity is limited to distance scales that are much shorter than 
distance scales that characterize gravitational considerations.  Any 
treatment of this topic that carries calculations to infinity, cannot be a 
valid treatment.  The treatment suggested by Leibovitz and Peres [4], 
considers a system which is neutral.   

In a recently published paper,  Rohrlich[3] discusses this topic, and  
 criticizing earlier works (including his earlier work[7]),
 he considers a treatment that should solve the 
problem.  Let us discuss  here briefly the treatment suggested in [3].     

The equation of motion given in [3] is: 

$$ m_0\dot{v}^{\mu} = F^{\mu}_{ext} + F^{\mu}_{self}       \eqno(5)  $$ 
\no where $m_0$ is the physical rest mass involved in the process ($m_0 = m+\delta m$, 
 [3]), and  $F^{\mu}_{self}$  actually equals  Abraham 4-vector. 
$\delta m$  represents the electromagnetic  mass which will be discussed 
 later.   Rohrlich  
 obtained this equation by generalizing the equation of motion, and by demanding that 
the constant in the generalized equation be ${2e^2\over 3c^3}$. 
 Rohrlich concludes that the vanishing of Abraham 4-vector 
($F^{\mu}_{self}$ in his notation) shows that the radiation energy comes 
 from the Schott term.  

This argumentation makes sense but it 
 raises a question:  The terms that appear in Abraham 4-vector 
in this case are the zeroth components of a 4-vector (components that represent 
power, and not force, see [3]).  
The question where is the force that does the work is still 
there.   We are looking for a three dimensional force that acts as a reaction force,  
 and we do not find it in this 
treatment.   The Schott term in this case represents the power performed by the 
force we are looking for, but not the force itself.   
 We discuss this point in the next section. 

\bg

\no {\bf  4. The stress force solution }  

\bg  

When an electric charge is accelerated, its electric field is not accelerated 
with the charge.  It is detached from the charge, and remains inertial.  Hence the 
electric field of the charge becomes curved[8].  
Fulton and Rohrlich [5] calculated the electromagnetic fields of a 
charge moving in a hyperbolic motion, along the $z$ axis, using the retarded 
potentials method.  The equation for the electric field lines were calculated 
by Singal[6],  and they are drawn in Fig. 2.  

\begin{figure}
\centering\epsfig{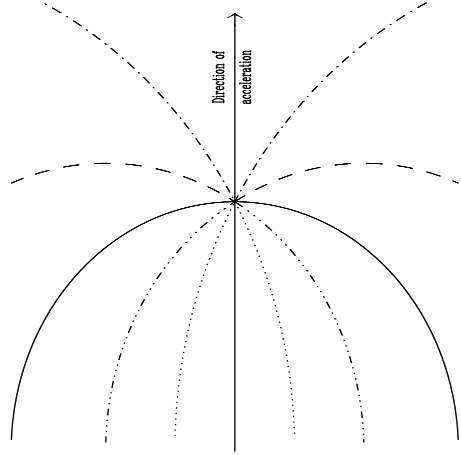} 
\vskip 0.2cm
\caption{{\small A curved electric field  of a uniformlly accelerated charge.}}
\end{figure} 

 The same results  were obtained 
by Gupta and  Padmanabhan[9], where they calculate the electric fields 
of an accelerated charge in the rest system of the charge, and then transform 
them to the inertial frame. They also show that by using the retarded 
coordinates ($x_{ret}, t_{ret}$) for the accelerated charge, one obtains 
the expressions for the electromagnetic fields as given in the 
standard textbooks[2,10].

The radius of curvature of the field lines is: 
  $R_c = c^2/a\sin\theta$,  
where $\theta$ is the angle between the initial direction of the field 
line and the acceleration.  Now, we follow in brief the calculations 
given in [11].   
In a curved electric field a stress force exists,  
 whose  density, $f_s$, is given by:  
 
$$ f_s = {E^2\over 4 \pi R_c},    \eqno(6) $$ 
\no where $E$ is the electric field, and $R_c$ is the radius of 
curvature of the field lines.  
 In the immediate vicinity of the charge, the electric 
field can be taken as  $E = e/r^2$, and we have: 

$$ f_s = {E^2\over 4\pi R_c} = {a \sin \theta\over 4\pi c^2} 
{e^2\over r^4}  \eqno(7)  $$ 

 The stress force is perpendicular to the field lines, so 
that the component of this force along the acceleration is $-f_s \sin 
\phi$, where $\phi$ is the angle between the local field line 
and the acceleration.  In the immediate vicinity of the charge 
 ($r \ll c^2/a$, where the direction 
of the field line did  not change much from its initial direction),  
 $\phi \sim \theta$, and we can write for  the parallel component 
of the stress force:

$$ -f_s \sin\phi \simeq -f_s \sin\theta = -{a\sin^2\theta\over 4\pi c^2} 
  {e^2\over r^4}  \eqno(8) $$ 

It is interesting to note that the angular distribution of 
the parallel component of the stress force is similar 
to that of the radiation.  
We have to sum over the parallel component of the stress force, $f_s$,    
and calculate the work done against this force.   

In order to sum over $f_s$,   we have to 
integrate over a sphere whose center is located on the charge.  
Naturally, such an integration involves a divergence (at the center). 
To avoid such a divergence, we take as the lower limit of the 
integration a small distance from the center, $r = c \Delta t$, (where 
$\Delta t$ is infinitesimal).  
  We calculate the work done by the stress 
force in the volume defined by  $c \Delta t \leq  r \leq r_{up}$, 
where $r_{up}$ is some large distance from  the charge, satisfying 
the demand:  $c \Delta t \ll r_{up} \ll c^2/a $.   
These calculations are performed in a system of reference $S$, 
which is a flat system that  momentarily coincides with the frame of 
reference of the accelerated charge  
at time $t = 0$, at the charge location.   

 Since the calculations are performed in the flat system  $S$,  
 the integration 
can be carried without using any terms concerning space curvature.  
Integration of the stress force  over a volume   
 extending from $r=c \Delta t$ 
to $r_ {\rm up}$, 
yields the total force due to stress, $F_s$:   

$$  F_s = 2 \pi  \int _{c \Delta t}^{r_{\rm up}} r^2 dr 
\int_0^{\pi} \sin \theta d \theta [- f_s \sin \theta ]
= - {{2}\over{3}} {{a}\over{c^2}} {{e^2}\over{c \Delta t}}
\left(1-{{c \Delta t}\over{r_{\rm up}}} \right) . \eqno(9)  $$
 
 Clearly the second term in the parenthesis can be neglected.  
$F_s$ is the reaction force.    
The power supplied by the external force on acting against the
electric stress is $P_s= - F_s  v $, 
where  $v$ is the velocity of the charge in system $S$ at time 
 $t = \Delta t$, $v = a \Delta t$. Substituting this value for $v$ 
we get for $P_s$:  

$$ P_s = {{2}\over{3}} {{a^2 e^2}\over{c^3}} . \eqno(10)  $$  

\no  This is the power radiated by an accelerated charged particle 
 at zero velocity (eq. 3). 

In order to include this result in the equation of motion (eq. 5), we 
should add the force  $-F_s$ to the external force,  which is the force 
needed to overcome the stress force, and we should include  $F_s$  in 
the spatial part of the Schott 4-vector (Schott term in the notation of [3]), 
 whose zeroth component is the power created by this force.   
The work performed by  $-F_s$ is the source of the energy carried by the radiation. 
  This extended equation includes all the forces involved in the 
process, including the force that  performs the work 
that creates the energy carried by the radiation. Rohrlich is right in 
his conclusion that the Schott term is the source of the power of the 
radiation,  but this definiftion becomes  complete when the stress force 
($F_s$ from eq. 9) is included in the spatial part of the Schott term.   

 The expression for the force in eq.  9 (before substituting the 
integration limits, $F_s = -{2\over 3}a e^2 
/c^2r$) equals the inertial force  ($4m_ea/3$) of the electromagnetic mass 
of the charge as calculated by Lorentz (see ref. [2], p. 790).  However, a 
work performed against an inertial force cannot be the  work that creates the 
energy carried by the radiation, because such a work creates a kinetic 
 energy of the electromagnetic mass  $m_e$, and this leaves us with the 
energy balance paradox.   Actually, this electromagnetic mass is already 
included in the equation of motion by Rohrlich, (eq. 5) as $\delta m$ (see [3]), 
and the force needed to accelerate this mass is already included in $F^{\mu}_{ext}$.

As an example, let us analyze the case of 
 an oscillatory motion of a charge in a linear antenna (length of $2D$), in the 
$x$ direction.   the equations of motion are:

$$ x = D\sin \omega t \ \ ; \ \ v = \omega D \cos \omega t  \eqno(11.a) $$ 
$$ a = -\omega^2 D\sin \omega t = -\omega^2 x \ \ ; \ \
 \dot{a} = -\omega^3 D \cos \omega t = -\omega^2 v  \eqno(11.b) $$ 

\bg
At time $\Delta t$  (where $\Delta t$ is infinitesimal), we   use the approximations: 

$\sin \omega t \simeq \omega \Delta t$, and $\ \ \cos \omega t \simeq 1$:

$$ x =  \omega D \Delta t \ \ ; \ \ v = \omega D    \eqno(12.a) $$ 
$$ a = -\omega^3 D \Delta t   \ \ ; \ \   \dot{a} = -\omega^3 D    \eqno(12.b) $$ 

\no and we find that when we start from a point at which $a=0$, 
 $\ \ a_{(\Delta t)} = \dot{a} \Delta t$.  

The motion is linear and from eq. 9 we have the  reaction force  $F_{reac}$ 
created in a linear motion:  $F_{reac} = F_s = {-2 e^2\over 3 c^3}{a\over \Delta t}$.  

Substituting for $a$, we find:  $F_s = {-2 e^2\over 3 c^3}\dot{a}$,  which is exactly Schott 
term.   Jackson [2, eq. 17.8] calls this expression $F_{rad}$, relating it to the radiation 
reaction force.  Since we know that  in a linear motion, at low velocities $(\beta < 0.01)$, 
no radiation reaction force exists, we deduce that $F_{reac}$, that expresses the stress 
force in the curved electric field, should replace  jackson's $F_{rad}$,  
the 
non-existing radiation reaction force.

 \bg
\no{\bf  5. Non-Zero Velocity  }
\bg

 Equation (10) shows the power emitted by a uniformly 
accelerated charge calculated at zero velocity.  We can consider   
 the case of an accelerated charge, moving with low velocity. 
 From Fig. 1 we observe that in this case,  
 a part of the radiation is radiated forward, 
 and evidently, this part of the radiation imparts a backward momentum 
to the radiating charge, thus creating a  reaction force.   
 To calculate the power created by {\it this} reaction force, we follow 
the calculations given in [12].  We    
 multiply eq. 1  by $\cos \theta/c$, (to obtain the parallel 
component of the momentum flux of the 
 radiation), and integrate over the angles.  The 
integration yields for the parallel component of the
momentum flux   $(p_{par})$:  

$$p_{par} = {2\over 3}{e^2 (\gamma^3a)^2\over c^4} \beta , \eqno(13)$$   

\no  while the total absolute value of the momentum flux,  
 $p$,  is found  by dividing  eq. 2 by $c$.  
(The perpendicular momentum flux vanishes because of the 
symmetry in the plane perpendicular to the direction of motion, 
 but we still can compare the parallel 
component of the momentum flux to the total absolute value of the 
momentum flux of the radiation).   By dividing $p_{par}$ by $p$,  
we find the weight of the parallel component of the momentum flux in 
the total absolute value of the momentum flux:

$${p_{par}\over p} =  \beta     \eqno(14) $$

This fraction is the weight of the parallel component of the 
momentum flux of the radiation, and this fraction creates a reaction
force - a radiation reaction force. 
 To get the weight of the work done by the radiation reaction 
force we should multiply this fraction by $\beta c$ (the velocity 
of the charge in the rest frame), and we find that 
 the weight of the work 
done in overcoming  the radiation reaction 
force (the reaction force created by the radiation) 
 in the total power radiated is  $\beta^2$.   The other part of the 
energy $( 1 - \beta^2 = 1/\gamma^2) $, is created by the stress force that 
 exists in the curved electric field.    
We find that the relative weight of the work done by the stress force in the total 
work done by the total reaction force decreases when $\gamma$ increases.  

\bg

\no {\bf  5. Conclusions }  

 \bg

The solution to the energy balance paradox is found when we consider the 
stress force that exists in the curved electric field of an accelerated 
charge.  At low (zero) velocities, 
 the work done by the external (accelerating) force in overcoming 
the stress force is the source of the energy carried by the radiation.  
The stress force is the spatial component of the Schott  4-vector, which is 
usually called the Schott term in Abraham 4-vector.  Thus,  the radiation 
 energy comes  from the Schott term, as suggested by Rohrlich[3].  
 
% \end{document}  

\bg

 \vfil
  \eject 
\bg
                                                         
\no {\bf References}
\bg
 
\no 1. Rindler W., {\it Special Relativity}, Sec. Ed., Oliver and Boyd, 
 New York, (1966).    

%\no 3. Landau L., Lifschiz E.M., {\it Classical Theory of Fields}, third Ed.,  
%Pergamon Press, Oxford, pp. 43 (1971)   

\no 2. Jackson J.D., {\it Classical Electrodynamics}, Sec. Ed., John Wiley   
 and Sons, New York (1975).       

\no 3. Rohrlich F., ``The Self Energy and Radiation Reaction", 
 {\it Amer. J. Phys.}, 68, (2000), 1109.    

\no 4. Leibovitz C., Peres A., ``The Energy Balance Paradox", 
 {\it Annals of Physics}, 25, (1963), 400. 

\no 5. Fulton R., Rohrlich F., ``Classical Radiation from a Uniformly Accelerated 
 Charge",  {\it Annals of Physics}, 9, (1960), 499.   

\no 6. Singal A. K., ``The Equivalence Principle and an Electric Charge in a 
Gravitational Field",  {\it Gen. Rel. Grav.}, 29, (1997), 1371. 

\no 7. Rohrlich F., {\it Classical Charged Particles}, Addison-Wessley, 
Reading, Mass., pp 111, (1965).  

\no 8. Harpaz A.,  ``The Nature of Fields",  {\it Euro. J.  Phys.}, 23, (2002), 263.   

\no  9. Gupta A., Padmanabhan T., ``Radiation from a Charged Particle and Radiation 
Reaction Reexamined",  {\it Phys. Rev.}, D57, (1998), 7241.

\no 10. Panofsky W.K.H., Phillips M., {\it Classical Electricty and Magnetism}, 
Sec. Ed., Addison Wesley Pub. Co., Reading MA, (1964).

\no 11. Harpaz A., Soker N., ``Radiation from a Uniformly Accelerated Charge", 
 {\it Gen. Rel. Grav.}, 30, (1998), 1217.   

\no 12. Harpaz A., Soker N., ``Radiation from an Electric Charge", 
 {\it Foundations of Physics}, 31, (2001), 935.   

\end{document}